# Dimensional hierarchy of topological bound states in the continuum

Shunda Yin[1†], Zhenyu Wang[1†], Liping Ye[1*], Hailong He[1], Manzhu Ke[1], Weiyin Deng[1*], Jiuyang Lu[1*], Zhengyou Liu[1,2*]

[1]Key Laboratory of Artificial Micro- and Nano-structures of Ministry of Education and School of Physics and Technology, Wuhan University, Wuhan 430072, China

[2]Institute for Advanced Studies, Wuhan University, Wuhan 430072, China

†These authors contributed equally to this work.

*Corresponding author. Emails: lpye@whu.edu.cn; dengwy@whu.edu.cn; jylu@whu.edu.cn; zyliu@whu.edu.cn

Bound states in the continuum (BICs), with the ability of trapping and manipulating waves within the radiation continuum, have gained significant attention for their potential applications in optics and acoustics. However, challenges arise in reducing wave leakage and noise from fabrication imperfections. The emergence of robust wave manipulations based on topological BICs (TBICs) offers promising solutions. Traditionally, TBICs of different dimensions are observed separately in distinct systems. Here, we report the experimental discovery of the coexistence of two-dimensional surface TBICs and one-dimensional hinge TBICs in a single three-dimensional phononic crystal system. Such an unprecedented dimensional hierarchy of TBICs is triggered by the mechanism of separability and protected by the valley Chern numbers. Notably, these TBICs inherit dispersive propagation characteristics from valley topology and can propagate robustly against defects without leakage. Our findings offer an efficient approach to multidimensional TBICs and can be applied in designing highly efficient acoustic devices for wave trapping and manipulation in multidimensional environments.

Bound states are studied in a wide range of systems, including both electronic and classical wave systems, and their existences may significantly modulate the physical properties of these systems and give rise to numerous intriguing phenomena [1-5]. BICs are localized states with energies lying in the continuum of extended states and are a general phenomenon arising from mechanisms such as wave interference, parameter tuning and separability [1]. Their unique properties of high Q-factors and strong localization enable BICs to facilitate advances in applications in sensing technologies [6-8], wave manipulation [9, 10], laser systems [11-13] and other domains [14, 15]. Topological boundary states, as another typical bound states, are localized states in topological materials whose energy lies in the band gap of the extended states [2-4], providing a plethora of interesting phenomena such as topological negative refraction/reflection [16, 17], topological lasers/sasers [18, 19] and quantum Hall effects [20-22]. Generally, topological boundary states are given by either the conventional bulk-boundary correspondence, or higher-order band topology [23]. Interestingly, higher-order band topology exhibits multidimensional topological physics, giving rise to the multidimensional topological boundary states [24-26]. For example, a three-dimensional (3D) second-order topological insulator exhibits the coexistence of two-dimensional (2D) topological surface states and one-dimensional (1D) topological hinge states.

Typically, in topological systems, extended states constitute a continuum spectrum and it is generally believed that no localized states can exist within this continuum due to hybridization. However, recent studies have demonstrated that topological boundary states can be embedded into the continuum of extended states without hybridization by combining BIC physics with band topology [27-38], forming the topological BICs (TBICs). These TBICs exhibit non-sensitivity to perturbations due to topological protection, which is quite different from conventional BICs that are easily perturbed/removed by perturbations such as parameter variations and sample defects [1, 14, 15, 39, 40]. To date, zero-dimensional (0D) corner and 1D edge/hinge TBICs have been widely realized [28-30, 32-38], where these TBICs in different dimensions are usually observed separately in systems of different dimensions. Notably, reports of 2D surface TBICs, which are theoretically predicted, are rare and have not been experimentally observed [27, 31]. This raises a natural question: can multidimensional TBICs coexist in a single system? In principle, a 3D topological material can support the coexistence of robust BICs of 2D surface, 1D hinge, and 0D corner states. However, it remains uncertain whether such 3D topological materials exist.

In this article, we design and fabricate a 3D phononic crystal (PC) that supports the coexistence of 2D surface and 1D hinge TBICs. Unlike previously reported TBICs and dimensional hierarchy of topological boundary states, which are derive from higher-order band topology, the dimensional hierarchy of TBICs revealed in our system originates from the first-order band topology, i.e., valley Chern numbers. We directly visualize the emergence of these 2D surface and 1D hinge TBICs and quantitatively measure their dispersions, manifesting a dimensional hierarchy of TBICs. Remarkably, our numerical simulations demonstrate that such multidimensional TBICs are robust against defects. The multidimensional TBICs considerably enrich the range of BICs and pave the way towards potential applications such as multidimensional energy harvesting and enhanced wave localization.

**Tight-binding model for multidimensional TBICs**

To illustrate the multidimensional TBICs, we start from a 3D tight-binding model constructed by stacking a 2D hexagonal lattice layer along the $z$ direction, as illustrated in Fig. 1(a). A unit cell includes five sites labeled $a$-$e$, which involves two identical sets of graphene sites ($a$, $b$ and $c$, $d$) connected indirectly through another site $e$. The Hamiltonian reads

$$H = \sum_{i,k} m e_{i,k}^\dagger e_{i,k} + \sum_{\langle ij \rangle, k} t_0 \left( a_{i,k}^\dagger b_{j,k} + c_{i,k}^\dagger d_{j,k} + \text{H.c.} \right)$$

$$+ \sum_{i,k} \left[ t_1 b_{i,k}^\dagger e_{i,k} + t_2 d_{i,k}^\dagger e_{i,k} + t_3 \left( b_{i,k}^\dagger e_{i,k+1} + d_{i,k}^\dagger e_{i,k+1} + b_{i,k}^\dagger e_{i,k-1} + d_{i,k}^\dagger e_{i,k-1} \right) + \text{H.c.} \right],$$

where $\delta_{i,k}$ ($\delta_{i,k}^\dagger$) with $\delta = a$-$e$ are the annihilation (creation) operators on the corresponding sublattice sites, $i$ denotes the position of the unit cell in each layer and $k$ represents the number of layers. $m$ is the on-site energy on site $e$, while $t_0$, $t_1$, and $t_2$ are the different intralayer couplings, and $t_3$ is the interlayer coupling. Its Bloch Hamiltonian $H_B(\boldsymbol{k})$ is obtained by the Fourier transformation, with $\boldsymbol{k} = (k_x, k_y, k_z)$ as the 3D Bloch wave vector. By applying a unitary transformation, $H_B(\boldsymbol{k})$ can be block-diagonalized to $H_D(\boldsymbol{k}) = H_2 \oplus H_3$, where $H_2 = \begin{pmatrix} 0 & h_{xy} \\ h_{xy}^* & 0 \end{pmatrix}$, $H_3 = \begin{pmatrix} 0 & h_{xy} & 0 \\ h_{xy}^* & 0 & h_z \\ 0 & h_z & m \end{pmatrix}$, and $h_{xy} = t_0 [1 + 2\cos(k_x/2) \exp(-i\sqrt{3} k_y/2)]$, $h_z = [t_1^2 + t_2^2 + 4(t_1 + t_2) t_3 \cos k_z + 8 t_3^2 \cos^2 k_z]^{1/2}$, see more details in Ref. [41]. All lattice constants are assumed to be 1 for simplicity. As a result, such a system can be decomposed into two independent subsystems with Hamiltonians $H_2$ and $H_3$, where these two subsystems are denoted as $h^{(2)}$ and $h^{(3)}$, respectively. Note that the eigenstates of these two subsystems are orthogonal, which facilitates the construction of bound states in the continuum.

The 3D Brillouin zone (BZ) and its surface projections for the lattice model are presented in Fig. 1(b). The bulk band dispersions are shown in Fig. 1(c), where the gray and orange curves denote the bands of $h^{(2)}$ and $h^{(3)}$, respectively. One can see that there is no band gap for $h^{(2)}$, but two band gaps for $h^{(3)}$. The topological properties of the upper gap for $h^{(3)}$ can be described by the $k_z$-dependent valley Chern number $C_{KH}^{(3)}(k_z)$, which is equal to $1/2$ along the $KH$ line [41], indicating that $h^{(3)}$ is a 3D valley topological insulator. This can result in the topological interface states and form the 2D TBICs. Moreover, the system possesses $k_z$-directional nonzero Zak phases $\theta_Z(k_x, k_y)$ which are calculated by the unit cell of $t_1$ as the interlayer coupling and close to $\pi$ [41], giving rise to the boundary states on the top surface, as shown by the black curves in Fig. 1(d) (The detailed top and bottom surface truncations are specified in the right panel). Interestingly, the bands of the top surface states host a 2D valley topological insulator phase with valley Chern number $C_{\bar{K}} = 1/2$ [41], and can lead to the topological hinge states and generate the 1D TBICs.

To exhibit the multidimensional TBICs, besides the above phase I, we construct the phase II by mirroring the phase I, which has the same bulk and top surface band dispersions but the opposite valley topologies as phase I [41]. When these two phases are put together to form an interface, as shown in the right panel of Fig. 1(e), the 3D valley topology of $h^{(3)}$ gives rise to the interface states. As shown in the left panel of Fig. 1(e), the dispersion of topological interface states (green curve) emerges in the upper gap of $h^{(3)}$. More importantly, these topological interface states are embedded in the continuum of the bulk states of $h^{(2)}$ (gray shadows) in both the $k_x$ and $k_z$ directions. Due to the decoupling of the $h^{(2)}$ and $h^{(3)}$ subsystems from each other, hybridization between these interface and bulk states is forbidden, leading to 2D interface TBICs. Meanwhile, the hinge modes emerge between the top surfaces of phases I and II, as shown in the right panel of Fig. 1(f), due to the 2D valley topology of the top surfaces. We calculate the hinge-projected dispersions of a rectangular-shaped structure along the $k_x$ direction, as shown in the left panel of Fig. 1(f). As we can see, besides the interface TBICs (green dots), topological hinge states (red curve) exist in the band gap of the top surface states (black dots). The hinge states are also embedded in the bulk states of $h^{(2)}$, forming the hinge TBICs. Therefore, our system possesses the dimensional hierarchy of TBICs.

**Acoustic realization of the tight-binding model**

The tight-binding model can be directly implemented in a PC of cavity-tube structure filled with air, as shown in Fig. 2(a). Specifically, as sketched in the inset of Fig. 2(a), a unit cell (take phase I as example) consists of five geometrically identical cylindrical cavities (of diameter 9.35 mm and height 19.5 mm) connected by tubes, where the cylindrical cavities emulate sites $a$-$e$ and the tubes introduce couplings $t_0$-$t_3$ between them. The special on-site energy of site $e$ is realized by attaching a small cuboid cavity (of side length 4.95 mm and height 1.8 mm) to the top and bottom of cylindrical cavity $e$ (colored in pink). The lattice constants in the $x$-$y$ plane and the $z$ direction are $a_0 = 54$ mm and $h = 27.9$ mm, respectively, and other parameters can be found in Ref. [41]. Similarly, the unit cell of phase II can be constructed by mirroring phase I.

Figure 2(b) presents the simulated bulk band dispersions of phase I for the dipole modes, where gray and orange curves denote the bands from $h^{(2)}$ and $h^{(3)}$ subsystems, respectively. It is found that two double band degenerations appear at points $K$ and $H$ for $h^{(2)}$ subsystem, and two complete band gaps exist in the bands of $h^{(3)}$ subsystem with the operating gap of 9.50-9.75 kHz. We then consider a phase I PC ribbon with the same surface truncations as those shown in the right panel of Fig. 1(d) to demonstrate the top surface states. The simulated dispersion is shown in the left panel of Fig. 2(c), where the black curves are the top surface states. The eigenfield of surface state at $\bar{K}$ point is presented in the right panel, which is highly localized near the top surface and decays in the bulk. In experiment, a broadband point sound source is fixed at the center of the top surface of phase I to excite the surface states, and a microphone is inserted into the surface cavities to probe the sound signals. The measured projected dispersions are displayed visually with bright color in Fig. 2(d), which are obtained from 1D Fourier transforming the measured pressure field, agreeing well with the simulated ones (white dots). We have checked that the PC of phase II has the identical band dispersions but opposite band topologies with that of phase I.

**Acoustic TBICs in a hierarchy of dimensions**

In the following, we numerically and experimentally validate the dimensional hierarchy of TBICs in PC. The projected dispersions of a PC ribbon with the XZ interface between phases I and II are calculated in the left panel of Fig. 3(a). Green curve denotes the topological interface states. It embeds in the continuum of the projected bulk bands of $h^{(2)}$, as the 2D TBICs. The right panel of Fig. 3(a) shows the strong interface localization of the 2D TBICs. These interface TBICs can be identified by our airborne sound experiments. To excite them, we use two sets of anti-phase point sources with $\pi$ phase shift, as shown in the inset of Fig. 3(b). The measured spatial acoustic pressure profile is presented in Fig. 3(b), where the acoustic signal is highly localized at the XZ interface and does not hybridize with the bulk, consistent with the simulated field in the right panel of Fig. 3(a). By 2D Fourier transforming the measured pressure field at the XZ interface, we obtain the isofrequency contours of the interface TBICs for a series of frequencies. As shown in Fig. 3(c), the experiment results (bright color) agree well with the simulations (green curves). The band broadening is due to the finite-size effect. We further extract the projected dispersions along the $k_x$ direction with $k_z = 0.5\pi/h$ and the $k_z$ direction with $k_x = 0.5\pi/a_0$, as depicted in Fig. 3(d). The measured interface state dispersions (bright color) are consistent with the simulated ones (green curves), and there is no signal of the overlapped bulk states. All these results explicitly demonstrate the existence of 2D interface TBICs.

Besides confirming the 2D interface TBICs, we further observe the 1D hinge TBICs on the hinge between the top surfaces of phases I and II. A pair of anti-phase sources with $\pi$ phase shift (denoted by the stars in Fig. 4(b)) are used to excite the hinge TBICs. As shown in the left panel of Fig. 4(a), the measured hinge state dispersions (bright color) capture the features of the simulated ones (red curve) perfectly and show the good excitation of the hinge TBICs. To further characterize the hinge TBICs, we also directly present the measured spatial acoustic pressure profile in Fig. 4(b). It shows a highly localized acoustic signal on the hinge (red line), in contrast to the nearly invisible acoustic signal in the bulk. This is consistent with the eigenfield of the hinge

TBICs shown in the right panel of Fig. 4(a). As a consequence, the aforementioned simulated and measured results on hinge and interface TBICs directly confirm the dimensional hierarchy of TBICs. In contrast to simply combining TBICs of different dimensions in spatially separated PCs [28-38, 41], dimensional hierarchical TBICs in a single 3D PC facilitate the development of integrated versatile acoustic devices and may inspire the next-generation technologies for communication and information processing.

**Conclusion and discussion**

In summary, we have presented a theoretical approach to constructing a dimensional hierarchy of TBICs, which have been validated by airborne sound experiments in a PC. The 2D interface and 1D hinge TBICs are protected by the 3D and surface valley topologies, respectively. These TBICs are embedded in the continuum of bulk states, yet remain highly localized at the interface or hinge boundaries without hybridization. All experimental data capture well with our theoretical predictions. More interesting, the surface polarization (second-order topology) gives rise to 0D corner TBICs, see more details in End Matter. Our system manifests the interplay of BIC physics and hybrid-order band topology. Different from traditional BICs, where practical Q-factors are often limited by sample disorders/defects [1, 14, 15, 39, 40], the multidimensional TBICs proposed here are demonstrated to be robust against sample imperfections (see End Matter), making high Q-factors achievable in practical applications. This work enables multidimensional topological boundary states in the continuum of bulk states, beyond the previous works on dimensional hierarchy of topological boundary states in isolated band gaps [25, 26]. Benefiting from the coexistence and integration of these multidimensional TBICs in a single material, our work facilitates topologically robust multidimensional wave trapping and manipulation even in the absence of spectral isolation, which may advance future communication and information technologies. For instance, surface and hinge TBICs can serve as robust waveguides, significantly enhancing the performance of subwavelength integrated photonics and phononic chips [8, 43-47] and injecting new vitality into the design of energy-efficient and high signal-to-noise ratio acoustic devices, such as acoustic sensors and energy harvesting devices. Moreover, one can also combine TBIC with other physics, such as nonlinear [34], non-Hermitian [15], and non-local effects [48-50], to inspire more interesting phenomena and applications.

**Acknowledgements**

This work is supported by the National Key R&D Program of China (Grants No. 2022YFA1404900, No. 2022YFA1404500, and No. 2023YFB2804701), the National Natural Science Foundation of China (Grants No. 12104347, No. 12222405, No. 12374409, and No. 12374419), and the Natural Science Foundation of Hubei Province of China (Grant No. 2023AFB604 and No. 2024AFB712).

**End Matter**

***0D corner TBIC in 3D lattice model and acoustic system.*** In this section, we discuss the 0D corner TBICs of our system. Theoretical analysis indicates that Wannier centers of the top surface states are located near the corners $B_0$ of each surface unit cell, see red stars in Fig. 5(a). Consequently, cutting along the boundaries marked by blue dashed lines in Fig. 5(a) results in the emergence of a 0D corner TBIC at top corner. Eigenvalue spectrum is shown in Fig. 5(b). As expected, one corner state (red sphere), is found in the band gap (yellow shadow) of top surface states and embedded in the spectrum of $h^{(2)}$ subsystem, forming the 0D corner TBICs. The eigenfield of the corner TBIC is shown in Fig. 5(c), showing strong localization at top corner. The 0D corner TBIC has been further numerically verified in the phononic crystal. Numerically calculated data are shown in Figs. 5(d)-5(f), in good agreement with the tight-binding model predictions.

***Robustness of the acoustic multidimensional TBICs.*** To characterize the topological robustness of multidimensional TBICs, we introduce valley-preserving defects and geometric size errors to the system. Valley-preserving defects are implemented by inserting some defect cavities, which are cylindrical cavities [depicted by the magenta color Figs. 6(c), 6(e) and 6(g)] with different geometrical parameters than the original cavities. Here, three cases of defect distributions are studied numerically: (i) defect cavities are inserted on the interface between phases I and II in the ribbon supercell [Fig. 6(c)], (ii) defect cavities are aligned in the *z*-direction of the interface centerline [Fig. 6(e)], (iii) defect cavities are arranged at random positions on the interface [Fig. 6(g)]. More details of the defect cavities can be found in Ref. [41]. We first calculate the band structures of the modified supercells, i.e., case (i), whose results are shown in Figs. 6(a) and 6(b). The band structures together with the corresponding eigenfields indicate that the topological interface and hinge states remain embedded in the bulk states of $h^{(2)}$ without hybridization, keeping the 2D interface and 1D hinge TBICs. We have also numerically investigated the transport properties of these multidimensional TBICs, as shown in Fig. 6(d). It can be seen that their transport behaviors remain the same as those in the unperturbed systems. Figures 6(f) and 6(h) present the simulated pressure fields of the multidimensional TBICs in the PCs with the defect distributions of cases (ii) and (iii), respectively. As can be seen from the figures, for both cases the acoustic waves propagate smoothly along the propagation paths and do not scatter into the bulks, despite suffering from defects. Moreover, to demonstrate that TBICs systems are robust to geometrical size errors, the systems with cavity diameter deviations in each unit cell i.e., lager $D_a$ and $D_c$, and smaller $D_b$, $D_d$, and $D_e$, have been constructed, as shown in Fig. 7(a). The corresponding band structures and eigenfields are presented in Fig. 7(b)-7(d). It can be observed that the topological interface and hinge states remain embedded in the bulk states of $h^{(2)}$ subsystem without hybridization, keeping the 2D interface and 1D hinge TBICs.

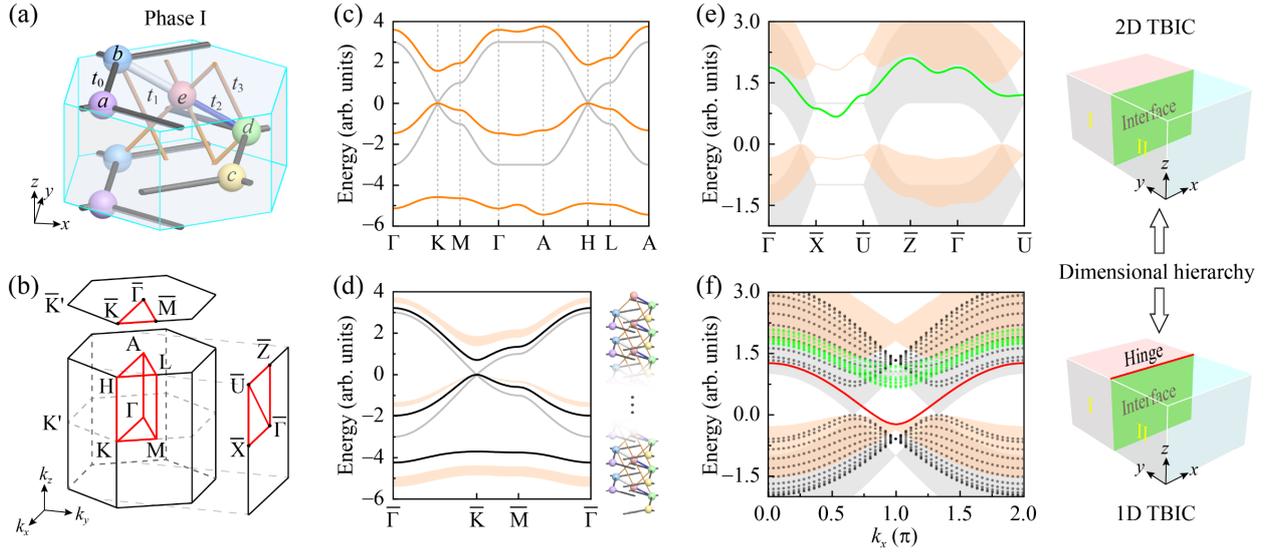

Fig. 1 Dimensional hierarchy of TBICs in a 3D lattice model. (a) Schematic of the lattice structure for phase I. A unit cell consists of five inequivalent sites labeled $a$-$e$. (b) The first BZ and its surface projections. (c) Bulk band dispersions along the high symmetry lines. Gray and orange curves denote the bands of $h^{(2)}$ and $h^{(3)}$ subsystems, respectively. (d) Projected dispersions (left panel) of a ribbon with the surfaces along the $z$ direction (right panel). Black curves represent the boundary states at the top surface. (e) Projected dispersions (left panel) of a ribbon with the interface normal to the $y$ direction (right panel). Green curve stands for the topological interface states which embed in the bulk states (gray shadows) and generate the 2D interface TBICs. (f) Projected dispersions (left panel) of a rectangular-shaped structure with period boundaries along the $x$ direction (right panel). Hinge states (red curve) embed in the bulk states (gray shadows) and form the 1D hinge TBICs. Gray and orange shadows in (d)-(f) represent the projected bulk states of $h^{(2)}$ and $h^{(3)}$ subsystems, respectively. The parameters are chosen as $t_0 = -1$, $t_1 = -2$, $t_2 = 1.5$, $t_3 = 0.5$, and $m = -3$.

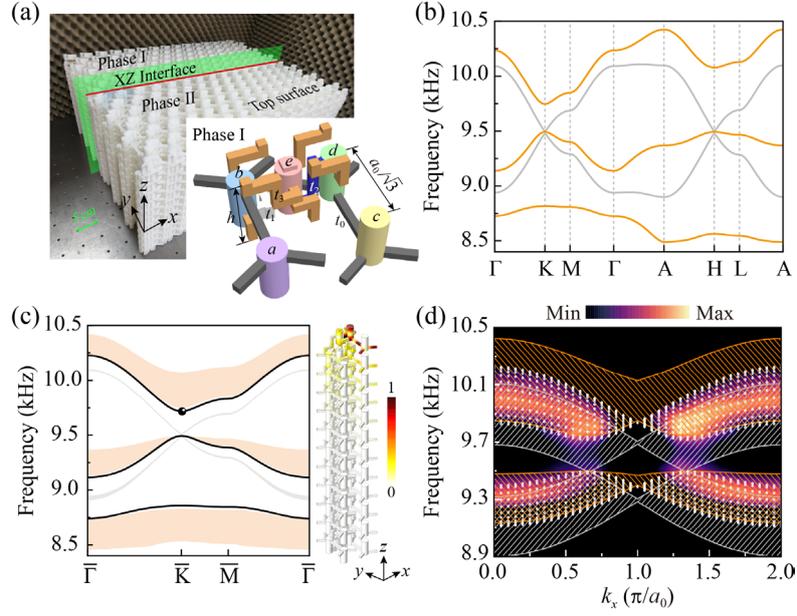

Fig. 2 Acoustic realization of the tight-binding model. (a) Photograph of the 3D PC. Inset is the acoustic unit cell. (b) Bulk band dispersions. Gray and orange curves denote the bands of $h^{(2)}$ and $h^{(3)}$ subsystems, respectively. (c) Left panel: projected dispersions of a PC with the surfaces along the $z$ direction. Black curves denote the top surface states, while gray and orange shadows represent the projected bulk modes of $h^{(2)}$ and $h^{(3)}$. Right panel: acoustic pressure profile for the top surface state at the $\bar{K}$ point (black sphere). (d) Measured (color map) and simulated (white dots) projected dispersions of the top surface states along the $k_x$ direction. The gray and orange shadows are the projected bulk dispersions.

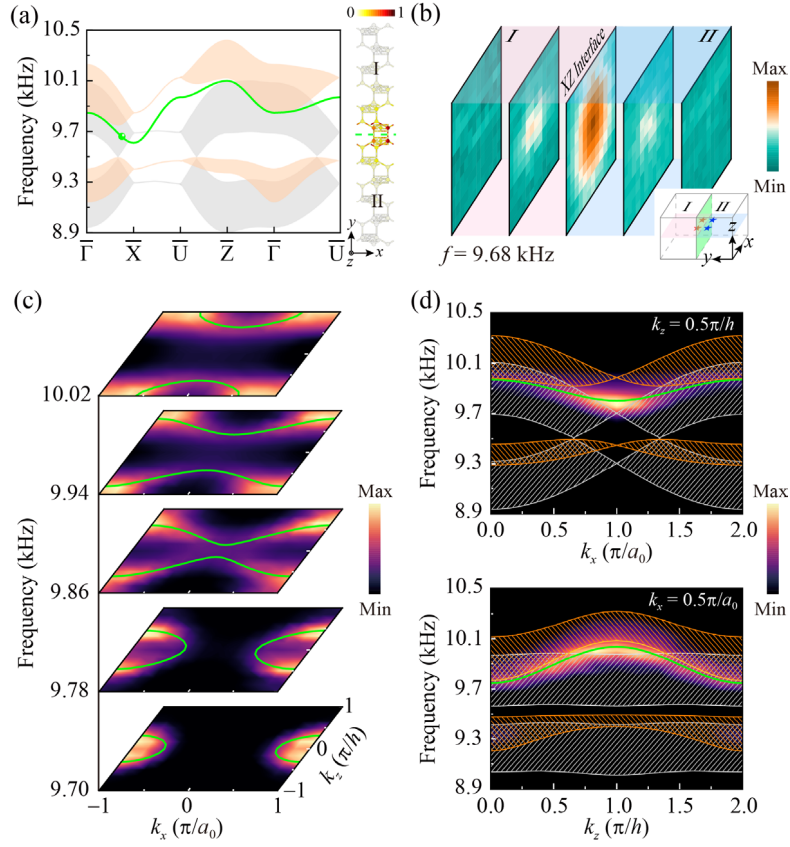

Fig. 3 Experimental observation of the 2D interface TBICs. (a) Left panel: Dispersion of the 2D interface TBICs (green curve) of the PC. Gray and orange shadows denote the projected bulk bands of $h^{(2)}$ and $h^{(3)}$, respectively. Right panel: acoustic pressure profile of the interface TBIC marked by the green sphere in dispersion. The green dashed line denotes the XZ interface. (b) Measured acoustic pressure profile of the interface TBICs at the operating frequency of 9.68 kHz. Inset shows the positions of two sets of anti-phase point sources (red and blue stars). (c) Measured (color map) and simulated (green curves) isofrequency contours of the interface TBICs. (d) Measured (color map) and simulated (green curves) projected dispersions of the interface TBICs along the $k_x$ direction with $k_z = 0.5\pi/h$ (top) and the $k_z$ direction with $k_x = 0.5\pi/a_0$ (bottom).

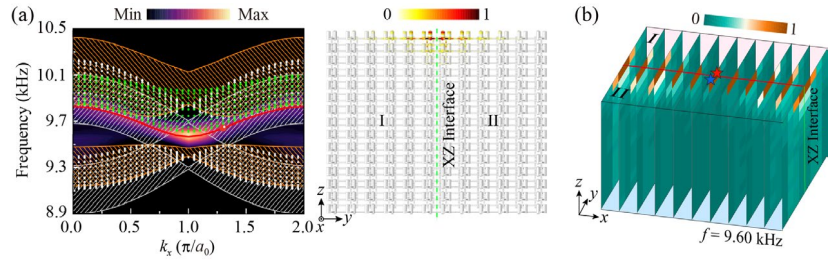

Fig. 4 Experimental observation of the 1D hinge TBICs. (a) Left panel: measured (color map) and simulated (red curve) dispersions of the hinge TBICs of the PC. Green and white dots denote the simulated XZ interface states and top surface states. Gray and orange shadows represent the projected bulk bands of $h^{(2)}$ and $h^{(3)}$, respectively. Right panel: acoustic pressure profile of the hinge TBICs marked by the red sphere in the left panel. (b) Measured acoustic pressure profile of the hinge TBICs at the operating frequency of 9.60 kHz. The red and blue stars denote the two anti-phase point sources, and the red line marks the hinge.

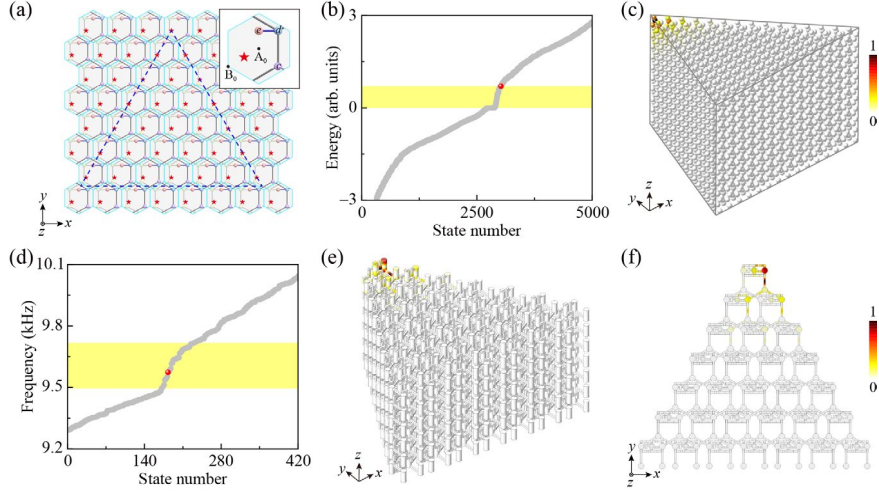

Fig. 5 0D corner TBIC in 3D lattice model and acoustic system. (a) Top view of the 3D lattice model, with red stars denoting the Wannier centers of the top surface states. Inset: unit cell of the top surface which includes three sites (denoted by $c$, $d$, and $e$). $A_0$ and $B_0$ denote the center and a corner of the surface unit cell, and the Wannier center (red star) of the top surface state is located at the center of $A_0B_0$ line in each unit cell. (b) Calculated eigenvalue spectrum for the triangular prism formed by cutting the 3D lattice along the blue dashed lines in (a). Red sphere represents a corner state which embeds in the bulk states (gray dots) and generates the 0D corner TBIC. Yellow shadow denotes the band gap of the top surface states. (c) Eigenfield of the 0D corner TBIC marked by the red sphere in (b). (d) The same as (b), but for acoustic system. (e), (f) Side and top views of acoustic pressure profile for the 0D corner TBIC marked by the red sphere in (d), respectively.

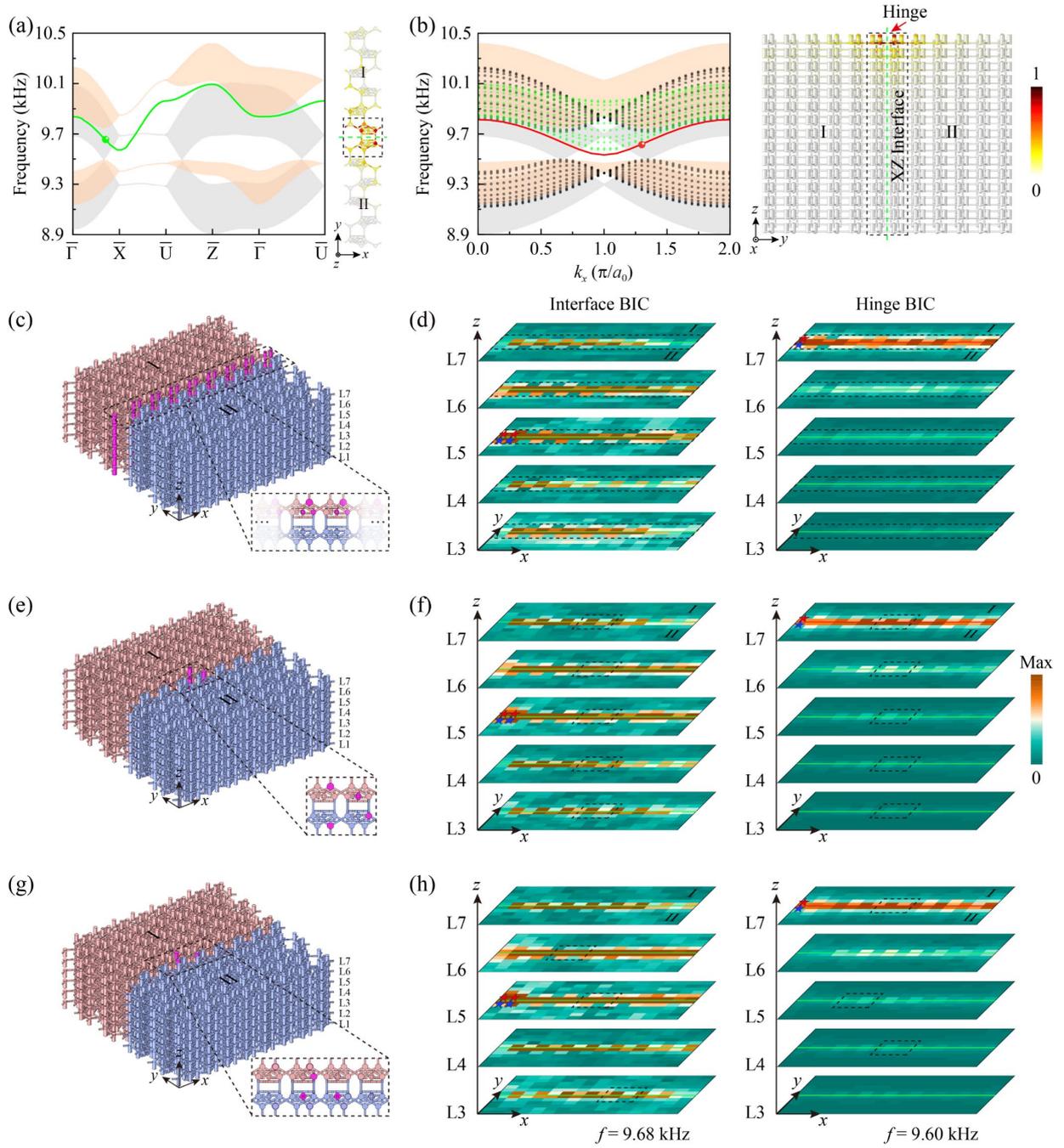

Fig. 6 Robustness of the acoustic multidimensional TBICs against valley-preserving defects. (a) Left panel: dispersion of the 2D interface TBICs (green curve) of the PC supercell with defects of case (i). Gray and orange shadows denote the projected bulk bands of $h^{(2)}$ and $h^{(3)}$, respectively. Right panel: acoustic pressure profile of the interface TBICs marked by the green sphere in the left panel. The green dashed line denotes the XZ interface. The black dashed box marks the position of the defect. (b) The same to (a), but for the hinge TBICs (red curve). Green and black dots in the dispersion denote the simulated XZ interface states and top surface states, respectively. (c) Phononic crystal with defects (magenta) of case (i). (d) Simulated acoustic pressure profiles of the interface TBIC at 9.68 kHz and hinge TBIC at 9.60 kHz for the structure in (c). The red and blue stars denote the two anti-phase point sources. The green and red line mark the XZ interface and hinge, respectively. (e) and (f), (g) and (h) are the same to (c) and (d), but for the structure with defects of cases (ii) and (iii), respectively.

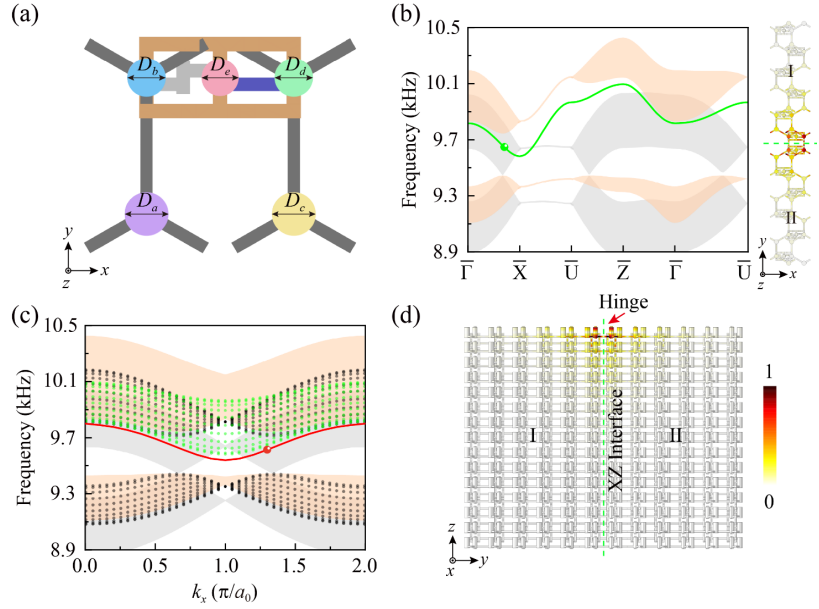

Fig. 7 Robustness of the acoustic multidimensional TBICs against geometric size errors. (a) Top view of the unit cell with cavity size deviations. Compared to the cavities used in the main text, cavities $a$ and $c$ have larger diameters $D_a = D_c = 10.0$ mm, while cavities $b$, $d$ and $e$ have smaller sizes $D_b = D_d = 8.8$ mm, $D_e = 8.5$ mm. (b) Left panel: dispersion of the 2D interface TBICs (green curve) of the PC supercell composed with the unit cell in (a). Gray and orange shadows denote the projected bulk bands of $h^{(2)}$ and $h^{(3)}$ subsystems, respectively. Right panel: acoustic pressure profile of the interface TBIC marked by the green sphere in the left panel. The green dashed line denotes the XZ interface. (c) The same to (b), but for the hinge TBICs (red curve). Green and black dots denote the XZ interface states and top surface states, respectively. (d) Acoustic pressure profile of the hinge TBIC marked by the red sphere in (c).